\documentclass[aps,10pt,twocolumn,superscriptaddress,floatfix]{revtex4-2}
\pdfoutput=1

\usepackage{amsmath}
\usepackage{amssymb}
\usepackage{amsthm}
\usepackage{bm}
\usepackage{enumitem}   % needed for \begin{enumerate}[label=(\roman*)] in Definition 1
\usepackage{physics}    % provides \ket, \bra, \Tr
\usepackage[colorlinks=true,linkcolor=blue,citecolor=blue,urlcolor=blue]{hyperref}

\theoremstyle{plain}
\newtheorem{theorem}{Theorem}
\newtheorem{proposition}{Proposition}
\theoremstyle{definition}
\newtheorem{definition}{Definition}

\begin{document}

\title{No generalised contextuality from the vacuum}

\author{Ruben Campos Delgado}
\email{ruben.camposdelgado@itp.uni-hannover.de}
\author{Selman Ipek}
\email{selman.ipek@itp.uni-hannover.de}
\affiliation{Institut für Theoretische Physik, Leibniz Universität Hannover, 30167 Hannover, Germany}

\begin{abstract}
We show that, under certain conditions, generalised contextuality, {in the sense introduced by Spekkens,}
cannot be created from the vacuum of a quantum field.
Initial noncontextual qudit states do not become contextual just by interacting with a quantum field. This is verified in two scenarios. First, in a perturbative approach where the time evolution operator is expanded in powers of the coupling constant. Second, in the limit of zero energy gap in the spectrum of the free detector Hamiltonian. Our results are at the opposite side of what has been recently obtained in the literature regarding the harvesting of Kochen-Specker contextuality.
{Our results demonstrate that the existing notions of contextuality can be independent of one another and that contextuality harvesting depends on which precise notion is used.}
\end{abstract}

\maketitle

\section{Introduction}
The Kochen-Specker (KS) theorem shows that, for Hilbert spaces of dimension three or more, it is impossible to assign pre-defined values to an observable in a way that is independent of the measurement context, namely the set of operators which commute with the given observable \cite{Specker1960, kochen1967problem}. This form of contextuality is a property of observables alone. KS contextuality differs from the operational notion introduced by Spekkens, which asks whether the probabilities assigned to preparations and measurements admit a noncontextual ontological model \cite{spekkens2005contextuality}. A sufficient condition for a set of states to be noncontextual, with respect to all possible measurements, is for them to be linearly independent. The reverse is true only for pure states \cite{CamposDelgado:2025lyl}.

Recently, it was shown that contextuality, both in the KS definition and in the Spekkens generalisation, is an observer dependent notion when only the spin degrees of freedom of the system are taken into account \cite{CamposDelgado:2025acu, Hance:2026wfi}.
These results are rooted in the nature of relativistic quantum mechanics.
They are a consequence of the symmetries of the Poincaré group and the way they act on quantum states or observables. Nevertheless, the notion of contextuality within a full quantum field theory formalism has not been fully explored yet. 
In \cite{LeMaitre:2025fnt, Lima:2025ueu} it was shown that contextuality, in its sheaf-theoretic version \cite{AbramskyBrandenburger2011, AbramskyBarbosaMansfield2017}, can be harvested by the vacuum state of a scalar field.

{In this paper, we follow a similar route, but instead ask whether \emph{Spekkens} contextuality can be created by interacting with a scalar field.}
We study this interaction by means of a Unruh-DeWitt detector, described, in the most general case, by $N$ qudits.
{This model has been used previously to study light-matter interactions~\cite{martin2013wavepacket,alhambra2014casimir}, as well as information-theoretic tasks, such as harvesting entanglement~\cite{pozas2015harvesting,perche2023harvesting} and `magic'~\cite{nystrom2025harvesting}.}
The main idea {of our work} is to derive an effective channel that describes the overall effect of the interaction of an observable with the field, after discarding the field degrees of freedom.
If the kernel of the channel is non-trivial, the channel maps linearly independent qudit states to linearly dependent ones. 
By the characterization of \cite{CamposDelgado:2025lyl}, there could {then} be a chance of creating generalised contextuality.
{Surprisingly, our result is negative:}
{for two important regimes,}
the effective channel always has a trivial kernel and no Spekkens contextuality can be created.
The {regimes} we consider are the perturbative expansion of the time evolution operator and the exact limit of zero energy gap between orthogonal energy eigenstates.
The paper is organized as follows. In Section \ref{sec:2} we review the notion of Spekkens (non)contextuality; Section \ref{sec:3} deals with a simple one-qubit detector;  Section \ref{sec:4} generalises to a $N$-qudit detector. 

\section{Spekkens contextuality}\label{sec:2}
We briefly review the notion of contextuality due to Spekkens \cite{spekkens2005contextuality, CamposDelgado:2025lyl}.  
Consider a theory where the primitive elements are preparations and measurements.
Let $\{P\}$ be the convex set of preparations and $\{M\}$ the convex set of measurements, with outcomes labelled by $k$.
{An ontological model reproduces the operational probabilities according to:}
\begin{equation}\label{eq:basic_def}
    p(k|P,M) =\sum_{\tau} \mu_P(\tau)\zeta_{M,k}(\tau).
\end{equation}
Here, $\mu_P(\tau)$ are probabilities with $\sum_{\tau}\mu_P(\tau)=1$ for all $P$ and $\zeta_{M,k}(\tau)$ are response functions with $\sum_{k}\zeta_{M,k}(\tau)=1$ for all $M$ and $\tau$.
If both $\mu_P$ and $\zeta_{M,k}$  depend only on the operational equivalence classes, i.e. classes of preparations (respectively measurements) that are not distinguishable by any measurements (respectively states), then the operational theory is said to be non-contextual.
Moreover, $\mu_P(\tau)$ and $\zeta_{M,k}(\tau)$ are linear functions of the equivalence classes of preparation procedures and measurement procedures, respectively.

In finite-dimensional quantum theory, a preparation equivalence class is a quantum state $\rho$, namely a positive semi-definite unit-trace linear operator acting on a finite-dimensional Hilbert space. Equivalence classes of measurements are described by positive operator valued measures (POVM), namely collections of positive semi-definite linear operators $M=\{ M_k\}_k$ which sum to the identity $\sum_k M_k=I$. Hence, noncontextuality in quantum theory can be defined as follows. 

\begin{definition}\label{def1}
The set of states $\{\rho_s\}_{s}$ is Spekkens non-contextual if for any POVM $\{M_k\}_k$, the probability of observing outcome $k$ can be expressed as
\begin{equation}\label{eq:spekken_def}
\Tr(\rho_s M_k) = \sum_{\tau \in \mathcal T} \Tr(\rho_s G_{\tau})\Tr(\sigma_\tau M_k) \quad \forall s,\,\, \forall M_k,
\end{equation}
where $\tau \in \mathcal T$ is an index set, $\Tr(\rho_s G_{\tau})\geq 0$, $\Tr(\sigma_{\tau} M_k)\geq 0$, and $\Tr(\sigma_\tau) = 1\,\, \forall \tau$ and $\sum_{\tau} \Tr\left(\rho_s G_{\tau}\right) = 1\,\,\forall s$.
\end{definition}

{Otherwise, the set $\{\rho_s\}_s$ is \textit{contextual}.}
Spekkens noncontextuality can then be characterized by the linear independence of states \cite{CamposDelgado:2025lyl}:
\begin{theorem}
If the states $\{\rho_s\}_s$ are linearly independent, then they are non-contextual.
\end{theorem}

In preparation for later developments, let us denote the set of linear operators acting on the detector Hilbert space $\mathcal H$ by $ L(\mathcal H)$. A channel $\Lambda$ is a completely positive trace preserving linear map $\Lambda: L(\mathcal H)\to  L(\mathcal K)$, where $\mathcal H$ and $\mathcal K$ are finite-dimensional Hilbert spaces.
\begin{proposition}\label{pro:invertibility}
Let $\Lambda:L(\mathcal H) \to L(\mathcal H)$ be a channel that is invertible as a linear map. If the set $\{\rho_s\}_s$ is noncontextual as in Eq.~(\ref{eq:spekken_def}) then the set $\{\Lambda(\rho_s)\}_s$ is also noncontextual.
\end{proposition}
Notice that this result holds even when the initial set is linearly dependent.
The proof can be found in Appendix~\ref{sec:invertibility-proof}.

Next we ask whether interaction with a quantum field can transform an initially noncontextual family of states into a contextual one.
In particular, the output states must be linearly dependent, although linear dependence alone is not sufficient for contextuality.

\section{The one-qubit effective channel}\label{sec:3}
A generic Unruh-DeWitt detector can be modeled as a single qudit that moves through spacetime. We neglect curvature and restrict ourselves to Minkowski spacetime. The free Hamiltonian has the spectral decomposition
\begin{equation}
    H_D =
    \sum_{k=0}^{d-1}E_k \ket{k}\bra{k},
%E_0\ket{0}\bra{0}+E_1\ket{1}\bra{1}+\dots+E_{d-1}\ket{d-1}\bra{d-1},
\end{equation}
where $\{\ket{0},\dots\ \ket{d-1}\}$ is an orthonormal basis of $\mathcal{H}_d$, the $d$-dimensional Hilbert space of the qudit, and $E_i$, $i\in\{0,\dots ,d-1\}$ are the energy levels. 

Let $x=(t,\textbf{x})$ be a coordinate system for the comoving frame of the detector and consider a scalar field $\phi(x)$.
The detector probes the field by interacting with its vacuum $\ket \Omega$, or other physical field states, which belong to a Hilbert space $\mathcal{H}_\phi$.
The form of the interaction can be described by the Hamiltonian
\begin{equation}
    H_I(x) = \lambda \xi(x) \mathcal{O}(t) \otimes\phi(x).
\end{equation}
Here, $\lambda<<1$ is the coupling constant,  $\xi(x)$ is a smearing function that controls the region of spacetime where the detector interacts locally with the field, and $\mathcal{O}$ is an observable of the detector.
{For additional details on this setup, see e.g., \cite{perche2022localized}.}

In the interaction picture, the time-dependent operator $\mathcal{O}(t)$ is obtained via the free Hamiltonian of the system:
\begin{equation}
    \mathcal{O}(t) = e^{\frac{i}{\hbar}H_Dt} \mathcal{O} e^{-\frac{i}{\hbar}H_Dt}.
\end{equation}
A typical choice for $\mathcal{O}$ is 
\begin{equation}\label{eq:choice_O}
    \mathcal{O} = \ket{0}\bra{1}+\ket{1}\bra{2}+\dots+\ket{d-2}\bra{d-1}+\text{h.c.},
\end{equation}
{together with $E_m = mE$, with $m=0,\cdots,d-1$, for some constant $E$.}
The resulting model is known as a SU(2) detector \cite{Lima:2023pyt, Lima:2025ueu}. 

Let us compute the corresponding $\mathcal{O}(t)$. We first notice that, by expanding the exponential, $e^{iH_Dt/\hbar}=\sum_{k=0}^{d-1}e^{iE_kt/\hbar}\ket{k}\bra{k}$. 
Thus, using the orthonormality condition $\langle i\rvert j\rangle=\delta_{ij}$,
\begin{eqnarray}\label{eq:product}
%\begin{split}
e^{\frac{i}{\hbar}H t}\ket{i}\bra{j}e^{-\frac{i}{\hbar}Ht}
&=&
%\sum_{k=0}^{d-1}e^{\frac{i}{\hbar}E_kt}\ket{k}\langle k\rvert i\rangle\bra{j}\sum_{k'=0}^{d-1}e^{-\frac{i}{\hbar}E_{k'}t}\ket{k'}\bra{k'}\\
\sum_{k,k'=0}^{d-1}e^{\frac{i}{\hbar}(E_k-E_{k'})t}\ket{k}\langle k\rvert i\rangle\langle j\ket{k'}\bra{k'}\notag\\
&=&
e^{\frac{i}{\hbar}(E_i-E_j)t}\ket{i}\bra{j}.
%\end{split}
\end{eqnarray}
Inserting Eq. \eqref{eq:choice_O} into Eq. \eqref{eq:product}, {and using that $E_{k+1}-E_k = E$}, we immediately obtain
\begin{equation}\label{eq:O(t)}
    \mathcal{O}(t) = e^{-\frac{i}{\hbar}Et}\sum_{k=0}^{d-2}\ket{k}\bra{k+1}+\text{h.c.}
\end{equation}

%Let us denote the set of linear operators acting on the detector Hilbert space $\mathcal H_d$ by $\mathcal L(\mathcal H_d)$.
%Our goal is to compute the effective channel $\Lambda$ that maps an initial observable $A_i$ on $\mathcal{H}_d\otimes \mathcal{H}_{\phi}$ to a final observable $A_f$ on $\mathcal{H}_d$ as a result of the interaction generated by $H_I$ and after tracing out the field degrees of freedom.
{Our goal is to compute the effective channel
$\Lambda: L(\mathcal H_d)\to L(\mathcal H_d)$
that maps an initial operator $A_i$ on the detector Hilbert space $\mathcal{H}_d$ to a final operator $A_f$ also on $\mathcal{H}_d$ as a result of the interaction with the field.}
{More explicitly,
we tensor our initial $A_i$ with the field vacuum, turn on the field interaction generated by $H_I$, and then trace out the field degrees of freedom.}
%If the kernel of this is non-trivial it can map linearly independent states to linearly dependent ones, and in virtue of the characterization of \cite{CamposDelgado:2025lyl}, it can map noncontextual states to possibly contextual ones (with respect to all possible measurements).
{The induced channel $\Lambda$ is a linear map and if its kernel is non-trivial then it can map linearly independent states to linearly dependent ones.
By virtue of the characterization of \cite{CamposDelgado:2025lyl}, such a linear map could therefore potentially map noncontextual states to contextual ones, in the sense of Definition~\ref{def1}.
As observed in~\cite{CamposDelgado:2025lyl}, however, linear dependence does not directly imply that the states are contextual since these states could still be highly noisy, precluding any witness of contextuality.}

{We first derive $\Lambda$ for a single qubit, i.e. $d=2$.
In this case, $\Lambda$ can be represented as a matrix acting on $\mathbb{R}^4$.}
The reason behind this is that set of Pauli operators $\mathcal{B}_{\text{P}}=\{\sigma_i\}_{i=0,\dots,3}=\{I,X,Y,Z\}$ forms a basis of $\text{Herm}(\mathbb{C}^2)=\{A\in \text{Mat}_2(\mathbb{C}):A=A^{\dagger}\}$. This is a vector space of dimension 4 over $\mathbb{R}$ so $\text{Herm}(\mathbb{C}^2)\simeq \mathbb{R}^4$. Thus, we can represent an observable as a vector and $\Lambda$ as a $4\times4$ matrix. Given an observable $A$, the isomorphism maps it to $(c_0,\dots, c_3)$, where $c_i$ are the expansion coefficient of $A=\sum_i c_i \sigma_i$. By taking the trace of both sides and using $\Tr(\sigma_i\sigma_j)=2\delta_{ij}$, one can derive the $i$-th coefficient as $c_i =\frac{1}{2}\Tr(A\sigma_i)$. 

We perform the calculation in two distinct ways. First using perturbation theory, then using an exact Magnus expansion for the gapless case.
%In both cases we consider the simple example of the detector being a single qubit, i.e. we set $d=2$.
%The operator $\mathcal{O}(t)$ reduces to $\mathcal{O}(t)=e^{-iEt/\hbar}\ket{0}\bra{1}+e^{i E t/\hbar}\ket{1}\bra{0}$.
{In the case of a single qubit,} the operator $\mathcal{O}(t)$ reduces to $\mathcal{O}(t)=e^{-iEt/\hbar}\ket{0}\bra{1}+e^{i E t/\hbar}\ket{1}\bra{0}$.

\paragraph*{A perturbative expansion}
The time evolution operator can be expressed as a Dyson series:
\begin{equation}
U = \mathcal{T} \exp\left(-\frac{i}{\hbar} \int dx\,H_I(x) \right).
\end{equation}
Although $U$ in principle depends on time, i.e. $U=U(t)$, we will consider its asymptotic behaviour and take $t$ to be just an integration variable. 
We work up to second order in the coupling. The expansion is
\begin{equation}\label{eq:dyson}
    U = I + U^{(1)}+U^{(2)}+O(\lambda ^3),
\end{equation}
with
\begin{equation}
    U^{(1)} = -\frac{i}{\hbar} \int dx\, H_I(x),
\end{equation}
\begin{equation}
    U^{(2)}=-\frac{1}{\hbar^2}\int dx \int dx'\, \theta(t-t') H_I(x)H_I(x').
\end{equation}
%Let $\ket{\Omega}$ denote the vacuum state of $\phi$.
To illustrate the general idea, we pick at time $t=0$ the operator $\mathcal{X}=X\otimes\ket{\Omega}\bra{\Omega}$. 
After interacting with the field, $\mathcal{X}$ evolves to $U\mathcal{X}U^{\dagger}$.
In computing the partial trace $X'=\Tr_{\phi}(U\mathcal{X}U^{\dagger})$, we notice that the one-point function of $\phi$ vanishes~%
\footnote{One may as well consider centered quasi-free states $\ket{F}$, satisfying $\bra{F}\phi(x)\ket{F}=0$.}, i.e. $\langle \Omega \lvert \phi(x) \rvert \Omega \rangle = 0$.
This is due to the fact that in canonical quantization $\phi$ is expanded in terms of creation and annihilation operators, $\phi\sim a+ a^{\dagger}$ and $a\ket{\Omega} =0$.
In fact, all odd correlation functions vanish. At the end we obtain the effective map $(0,1,0,0)\to^{\hspace{-2.5mm}\vspace{2mm}\Lambda}(X_0,X_1,X_2,X_3)$, where $X_i = \frac{1}{2}\Tr(X'\sigma_i)$ contains the two-point function of $\phi$: $W(x,x') := \langle \Omega \lvert \phi(x)\phi(x') \rvert \Omega \rangle$.
The steps of the calculation are the same for the other three basis operators, each times tensored with $\ket{\Omega}\bra{\Omega}$.
With a natural choice of notation, the channel $\Lambda$ is then defined by the system of linear equations
\begin{equation}
\begin{cases}
&(I_0,I_1,I_2,I_3)^T=\Lambda (1,0,0,0)^T\\
&(X_0,X_1,X_2,X_3)^T=\Lambda (0,1,0,0)^T\\
&(Y_0,Y_1,Y_2,Y_3)^T=\Lambda (0,0,1,0)^T\\
&(Z_0,Z_1,Z_2,Z_3)^T=\Lambda (0,0,0,1)^T.\\
\end{cases}
\end{equation}
The solution has the structure $\Lambda = I + \lambda^2 \Lambda_2+O(\lambda^4)$. What is important to notice here is that the kernel of $\Lambda$ is trivial. In fact, at any fixed order $N$ in perturbation theory and for sufficiently weak coupling $\lambda$, $\Lambda$ is always invertible. To see this, expand $\Lambda$ and its (supposed) inverse as 
\begin{equation}
    \Lambda = \sum_{n=0}^N\lambda^n\Lambda_n+ O(\lambda^{(N+1)}), \hspace{2 mm}T = \sum_{n=0}^N\lambda^n T_n+ O(\lambda^{(N+1)}),
\end{equation}
with $\Lambda_0=T_0=I$. 
Their product is 
\begin{equation}
    \Lambda T = I+\lambda(T_1+\Lambda_1)+ \lambda^2(T_2+\Lambda_1T_1+\Lambda_2)+\dots
\end{equation}
which leads to a system of $N$ linear equations for $N$ unknowns (the $T_n$ matrices). The system has a unique solution: at order $n$, one inserts the expression for $T_n$ obtained by solving the corresponding equation at the $(n-1)$-th step. For example, at first order one finds $T_1=-\Lambda_1$; at second order $T_2 = \Lambda_1^2-\Lambda_2$, and so on. Hence $T=\Lambda^{-1} + O(\lambda^{(N+1)})$ and the channel never maps linearly independent states to linearly dependent ones.
This apparently innocent fact has an important consequence, namely that at finite order in perturbation theory Spekkens contextuality can never be created simply by interacting with a scalar field.
{In particular, this holds irrespective of the particular Unruh-DeWitt detector model that is considered.}

\paragraph*{The Magnus expansion for the gapless case} 
The operator in Eq. \eqref{eq:dyson} is a solution of the equation
\begin{equation}\label{eq:interaction_picture}
    i\hbar \frac{dU}{dt} = H_I(t)U,
\end{equation}
where 
\begin{equation}
    H_I(t) = \int d^3\textbf{x}\, H_I(t,\textbf{x}).
\end{equation}
Eq. \eqref{eq:interaction_picture} has the structure $Y'(t) = A(t) Y(t)$, $Y(t_0)=Y_0$, as soon as we identify $Y(t) = U(t)$, $A(t)=-\frac{i}{\hbar} H_I(t)$ Magnus \cite{Magnus1954} realised that the solution of such a differential equation can be written in terms of the exponential of an infinite number of terms:
\begin{equation}
Y(t)=e^{\Omega_1+\Omega_2+\dots}Y_0,  
\end{equation}
the first coefficients being
\begin{equation}
    \Omega_1(t)=\int^t dt' A(t')=-\frac{i}{\hbar}\int ^{t}dt'\,H_I(t'),
\end{equation}
\begin{equation}
\begin{split}
    &\Omega_2(t)= \frac{1}{2}\int ^tdt'\int^{t'}dt{''}\left[A(t'), A(t'')\right]=\\
    &-\frac{1}{2\hbar^2}\int ^ t dt'\int^t dt'' \theta(t'-t'') \left[H_I(t'),H_I(t'')\right] .
\end{split}
\end{equation}
Let us consider our one-qubit detector in the gapless case, namely where all the energy levels are the same: $E_i = E_j$, $\forall i,j$.
This {greatly} simplifies our calculations. First of all, $\mathcal{O}(t)$ becomes time-independent. From Eq. \eqref{eq:O(t)} it is $\mathcal{O}(t)=\ket{0}\bra{1}+\ket{1}\bra{0}=X$. The Magnus coefficients are
\begin{equation}
    \Omega_1 = -\frac{i}{\hbar}X\otimes\phi_{\xi},
\end{equation}
\begin{equation}
    \Omega_2 = -\frac{1}{2\hbar^2}I\otimes C_{\xi},
\end{equation}
where
\begin{equation}
    \phi_{\xi} = \lambda \int d^4x\, \xi(x)\phi(x),
\end{equation}
\begin{equation}
    C_{\xi} = \lambda^2 \int d^4 x \int d^4 x'\, \theta(t-t')\left[\phi(x),\phi(x')\right]\xi(x)\xi(x').
\end{equation}
Notice that $\left[\phi(x),\phi(x')\right]=i\Delta(x,x')$ is anti-hermitian, since it is the commutator of two hermitian operators. Thus, $C_{\xi}=i\tilde{C}_{\xi}$ is a purely imaginary number {times the identity} and the higher Magnus coefficients vanish, as they all involve nested commutators. The time evolution operator has the exact form
\begin{equation}\label{eq:U_qubit}
    U = e^{-\frac{i\tilde{C}_{\xi}}{2\hbar^2}}e^{-\frac{i}{\hbar}X\otimes \phi_\xi}.
\end{equation}
The first factor is an overall phase and is irrelevant for the following discussion. 
As before, we compute the transformed Pauli operators and trace out the field. Let us consider an initial operator $K\in \mathcal{B}_{\text{P}}$ tensored with $\ket{\Omega}\bra{\Omega}$. The effective channel maps it to 
\begin{equation}
    K' = \sum_{n,m=0}^{+\infty}\left(\frac{i}{\hbar}\right)^{n+m}\frac{(-1)^m}{n!m!}X^{m}KX^{n}\langle\Omega \lvert \phi_{\xi}^{n+m}\rvert \Omega \rangle.
\end{equation}
After applying the commutation properties of the Pauli matrices and resumming the exponentials, we find
\begin{equation}
\begin{split}
I\to I, \hspace{3mm}  X\to X,\\
Y\to a Y, \hspace{3mm} Z\to aZ,
\end{split}
\end{equation}
where
\begin{equation}
    a = \langle\Omega\rvert e^{2i\phi_{\xi}/\hbar }\rvert\Omega\rangle. 
\end{equation}
{For the smearing functions of interest, the smeared vacuum variance $\bra \Omega \phi_\xi^2\ket \Omega$ is finite so that the quantity $a$, defined above, is strictly positive.}
The corresponding channel is therefore diagonal with non-zero coefficients:
\begin{equation}
    \Lambda = 
    \begin{pmatrix}
      1 & 0 & 0 &0 \\
      0 & 1 & 0 & 0 \\
      0 & 0 & a & 0 \\
      0 & 0 & 0 & a
    \end{pmatrix}
\end{equation}
and thus has a trivial kernel. 
We conclude that the previous impossibility of generating Spekkens contextuality is not an artifact of the perturbative expansion. Rather, it appears even in presence of an exact expression for the time evolution operator, albeit in the special case of zero energy gap. In fact, the result still holds even for a small gap. The Magnus coefficients become in this case a series expansion in terms of the small gap parameter and the same argument of the previous section leads us to conclude that the effective channel is always invertible at any finite order.
{Alternatively, since the set of invertible matrices is open, invertibility of the gapless channel persists for sufficiently small energy gaps, provided the channel depends continuously on the gap.}

\section{The $\mathbf N$-qudit effective channel}\label{sec:4}
One may wonder whether the previous no-go result is a consequence of the simplicity of the detector, namely, it being a single qubit. In this section we prove that, in the gapless limit, the result still holds when the detector is made up of $N$ qudits. 

Let us first consider the case $N=1$. Again, the gapless limit is particularly simple because the detector operator becomes time-independent: $\mathcal{O}(t)=\mathcal{O}$. 
The time evolution operator is the generalisation of Eq. \eqref{eq:U_qubit}:
\begin{eqnarray}
U
&=&
e^{-\frac{i\tilde{C}_{\xi}\mathcal{O}^2}{2\hbar^2 }}e^{-\frac{i}{\hbar}\mathcal{O}\otimes \phi_{\xi}}\notag\\
&=&
\sum_j e^{-\frac{i\tilde{C}_{\xi}}{2\hbar^2}o_j^2}\ket{j}\bra{j}\otimes e^{-\frac{i}{\hbar}o_j\phi_{\xi}},\notag
\end{eqnarray}
where we have inserted the spectral decomposition $\mathcal{O}=\sum_j o_j \ket{j}\bra{j}$ and used $f(\mathcal{O})=\sum_j f(o_j)\ket{j}\bra{j}$.  Let us work in the basis made up of the outer products of eigenstates of $O$, $\mathcal{B}_O=\{\ket{i}\bra{j}\}_{ij}$, and consider the transformation of a generic $K=\ket{k}\bra{l}$. In this basis the channel is diagonal and acts as
\begin{equation}\label{eq:channel_qudit}
\Lambda(\ket{k}\bra{l})=\Lambda_{kl} \ket{k}\bra{l},  
\end{equation}
where the entries are
\begin{equation}\label{eq:elements_channelqudit}
    \Lambda_{kl}=e^{-\frac{i\tilde{C}_{\xi}}{2\hbar^2}\left(o^2_k-o^2_l\right)}\langle \Omega \rvert e^{\frac{i}{\hbar}\left(o_l-o_k\right)\phi_{\xi}}\rvert \Omega\rangle
\end{equation}
and they are all nonzero. 
Hence, the channel is always invertible for any $d$. In other words, a set of qudit states that is initially Spekkens non-contextual can not become Spekkens contextual just by interacting with a scalar field. 
As an aside, Eq. \eqref{eq:elements_channelqudit} recovers the qubit case by taking $\mathcal{B}_{O}=\{\ket{+}\bra{+},\ket{+}\bra{-},\ket{-}\bra{+},\ket{-}\bra{-}\}$ and $o_{\pm}=\pm 1$.

Let us now consider the most general case of $N$ qudits. The interaction Hamiltonian is
\begin{equation}
    H_I(x) = \lambda \sum_{\alpha=1}^N \xi_\alpha(x)\,\mathcal{O}_\alpha(t)\otimes \phi(x),
\end{equation}
where $\mathcal{O}_{\alpha}$ acts nontrivially on the $\alpha$-th tensor factor and trivially otherwise.
In the gapless limit $\mathcal{O}_\alpha(t)=\mathcal{O}_\alpha$ for all $\alpha$.
We define the smeared fields and the (real) smeared commutator coefficients
\begin{align}
    \phi_{\xi_{\alpha}} &= \lambda\int d^4x\,\xi_\alpha(x)\phi(x),\\
    \tilde{C}_{\xi_{\alpha\beta}} &= \lambda^2\int d^4x\int d^4x'\,\theta(t-t')\Delta(x,x')\xi_\alpha(x)\xi_\beta(x').
\end{align}
The first two Magnus coefficients are
\begin{align}
    \Omega_1 &= -\frac{i}{\hbar}\sum_{\alpha=1}^N \mathcal{O}_\alpha\otimes\phi_{\xi_{\alpha}},\\
    \Omega_2 &= -\frac{i}{2\hbar^2}\sum_{\alpha,\beta=1}^N \mathcal{O}_\alpha\mathcal{O}_\beta\otimes \tilde{C}_{\xi_{\alpha\beta}} .
\end{align}
Note that the cross terms $\alpha\neq\beta$ do \emph{not} cancel in general. 
Let $\ket{\bm{k}}=\ket{k_1}\otimes\cdots\otimes\ket{k_N}$ denote the joint eigenbasis of the $\mathcal{O}_\alpha$, with $\mathcal{O}_\alpha\ket{\bm{k}}=o_{k_{\alpha}}\ket{\bm{k}}$, and define
\begin{equation}
    \Phi_{\bm{k}}=\sum_{\alpha=1}^N o_{k_\alpha}\,\phi_{\xi_\alpha},
    \qquad
    \Theta_{\bm{k}}=\frac{1}{2\hbar^2}\sum_{\alpha,\beta=1}^N\tilde{C}_{\xi_{\alpha\beta}}\,o_{k_\alpha}o_{k_\beta}.
\end{equation}
The time evolution operator is then
\begin{equation}
    U=e^{\Omega_2}e^{\Omega_1}=\sum_{\bm{k}} e^{-i\Theta_{\bm{k}}}\ket{\bm{k}}\!\bra{\bm{k}}\otimes e^{-\frac{i}{\hbar}\Phi_{\bm{k}}}.
\end{equation}
Tracing out the field in the vacuum, the same calculation that led to Eq.~\eqref{eq:elements_channelqudit} gives a channel that is diagonal in the basis $\{\ket{\bm{k}}\!\bra{\bm{l}}\}$: $\Lambda(\ket{\bm{k}}\!\bra{\bm{l}})=\Lambda_{\bm{k}\bm{l}}\ket{\bm{k}}\!\bra{\bm{l}}$, with
\begin{equation}\label{eq:elements_channel_N}
    \Lambda_{\bm{k}\bm{l}}
    = e^{-i(\Theta_{\bm{k}}-\Theta_{\bm{l}})}\,
    \langle\Omega\rvert\, e^{\frac{i}{\hbar}\Phi_{\bm{l}}}\,e^{-\frac{i}{\hbar}\Phi_{\bm{k}}}\,\lvert\Omega\rangle .
\end{equation}
The commutator $[\Phi_{\bm{l}},\Phi_{\bm{k}}]=iD_{\bm{k}\bm{l}}$ is a purely imaginary c-number, so that $D_{\bm{k}\bm{l}}\in\mathbb{R}$. For a quasi-free (Gaussian) state such as the vacuum, using the Baker--Campbell--Hausdorff formula,
\begin{equation}
    \Lambda_{\bm{k}\bm{l}}
    = e^{-i(\Theta_{\bm{k}}-\Theta_{\bm{l}})+\frac{i}{2\hbar^2}D_{\bm{k}\bm{l}}}\,
    \exp\!\left[-\frac{1}{2\hbar^2}\langle\Omega\rvert(\Phi_{\bm{l}}-\Phi_{\bm{k}})^2\lvert\Omega\rangle\right].\notag
\end{equation}
The first factor is a pure phase. Hence $|\Lambda_{\bm{k}\bm{l}}|>0$ whenever the smeared vacuum variances are finite, and the channel is invertible. For $N=1$ one has $D_{kl}=0$ and Eq.~\eqref{eq:elements_channelqudit} is recovered.

One may wonder whether in the ordinary gapped case the kernel might become nontrivial. We cannot verify it explicitly since it would require an infinite number of Magnus coefficients. However, the Magnus expansion and the perturbative Dyson expansion give the same answer assuming that the latter converges, i.e. $\lambda<<1$. Therefore, by going to higher order in perturbation theory one would arbitrarily comes closer to the exact result of the Magnus expansion. Since we argued that at any order in perturbation theory the kernel is invertible, it must be invertible also in the general gapped case.

\section{Discussion}
We considered the Spekkens, or generalised, notion of contextuality and showed that initially noncontextual qudit states cannot become contextual by interacting with a quantum scalar field. We showed this in two distinct situations. Firstly, by truncating the Dyson series at a generic order in perturbation theory. Secondly, in the gapless case where all the orthogonal qudit states have the same energy. In both cases the effective channel that encodes the action of the field on the quantum states has a trivial kernel. Thus, this channel does not map linearly independent states to linearly dependent ones nor noncontextual states to contextual ones.
{Due to the general structure of our approach, we anticipate that these results would continue to hold in more general curved spacetimes, although in that case they would depend on the choice of vacuum, which is not typically unique for a generic spacetime.}

At first glance, our results appear to contrast with those obtained by LeMaitre \cite{LeMaitre:2025fnt} and Lima et al. \cite{Lima:2025ueu} who argued that Kochen-Specker contextuality, more specifically its sheaf theoretic formulation, can be harvested from the vacuum of a scalar field. It is worth stressing the fact that there is no contradiction. Indeed, there is no obvious relationship or arrow of implication between the different notions of contextuality~\cite{private}.
Our result is further evidence of this. Contextuality harvesting is not a fundamental phenomenon in relativistic quantum information. One has first to specify which type of contextuality one is referring to.

\begin{acknowledgements}
The authors would like to thank Philip LeMaitre and Martin Pl{\'a}vala for useful discussions.
S.I. acknowledges support from Nieders{\"{a}}chsisches Ministerium f{\"{u}}r Wissenschaft und Kultur and the Alexander von Humboldt Foundation.
\end{acknowledgements}

\bibliography{citations}

\appendix

\section{Invertible channels preserve non-contextuality}%
\label{sec:invertibility-proof}

We now prove Proposition~\ref{pro:invertibility}.
Note that the states $\{\rho_s\}_s$ need not be linearly independent.
Since the same noncontextual representation reproduces
the probabilities for all POVMs, Eq.~\eqref{eq:spekken_def} implies the operator identity
\[
\rho_s
=
\sum_\tau \Tr(\rho_s G_\tau)\sigma_\tau.
\]
This is a consequence of the fact that quantum effects span the Hermitian operator space.
Define now
\[
G'_\tau := (\Lambda^{-1})^*(G_\tau),
\qquad
\sigma'_\tau := \Lambda(\sigma_\tau),
\]
where $^{*}$ denotes the Hilbert--Schmidt adjoint.
For $\rho'_s := \Lambda(\rho_s)$, we have
\[
\Tr(\rho'_s G'_\tau)
=
\Tr\!\left(\Lambda^{-1}(\rho'_s)G_\tau\right)
=
\Tr(\rho_s G_\tau)
\geq 0,
\]
and therefore
\[
\sum_\tau \Tr(\rho'_s G'_\tau)=1.
\]
Moreover, each $\sigma'_\tau$ is a density operator, because $\Lambda$ is completely positive and trace preserving.
Applying $\Lambda$ to the initial decomposition gives
\[
\rho'_s
=
\sum_\tau
\Tr(\rho'_s G'_\tau)\sigma'_\tau.
\]
Consequently, for every POVM effect $E$,
\[
\Tr(\rho'_s E)
=
\sum_\tau
\Tr(\rho'_s G'_\tau)\Tr(\sigma'_\tau E).
\]
This is a noncontextual representation of the output family in the sense of Eq.~\eqref{eq:spekken_def}.

\onecolumngrid

\end{document}